# A Black Start Strategy for Hydrogen-integrated Renewable Grids with Energy Storage Systems

Jin Lu
Electrical & Computer Engineering
University of Houston
Houston, USA
jlu28@uh.edu

Linhan Fang
Electrical & Computer Engineering
University of Houston
Houston, USA
lfang7@uh.edu

Fan Jiang
Electrical & Computer Engineering
University of Houston
Houston, USA
fjiang6@uh.edu

Xingpeng Li
Electrical & Computer Engineering
University of Houston
Houston, USA
xli83@central.uh.edu

*Abstract*—With the increasing integration of renewable energy, the reliability and resilience of modern power systems are of vital significance. However, large-scale blackouts caused by natural disasters or equipment failures remain a significant threat, necessitating effective restoration strategies. This study proposes novel black start models for modern power systems that integrate fuel cells and battery storage, recognizing their distinct characteristics and contributions to grid resilience. These models specifically address the restoration of electrical grids, including the energization paths and time of the transmission network, while accounting for the unique power output traits of fuel cells and the energy storage capacity of batteries as black start resources. Black start simulations, comparing the generator startup sequence (GSUS) with fuel cell versus battery systems, are performed on the IEEE 39-bus system. We conduct sensitivity analyses on fuel cell capacity, battery storage capacity, initial state of charge (SOC), and resource locations to identify optimal scenarios for black start operations.

*Index Terms*-- Battery storage, Black start, Fuel cells, Grid restoration, Generator startup sequence.

## I. INTRODUCTION

With the development and expansion of the power system, grid reliability and resilience are of vital significance, especially considering the grid integration of fast-growing renewables. However, natural disasters and equipment failures can cause large-scale blackouts [1], which will loss electricity demand and hurt the reliability of power system. Hence, effective power system restoration strategy should be deployed to against outages and improve the power system resilient.

The restoration strategy prioritizes rapid load restoration while ensuring compliance with all power system operational constraints. In power system, restoration constitutes a complex optimization challenge characterized by multiple objectives, sequential stages, numerous control variables, and interdependent constraints [2]. Generally, it can be divided into three stages: 1) black-start (BS), 2) network reconfiguration, and 3) load restoration. During the BS phase, non-black-start (NBS) generators are restored using power from the BS generator. In the network reconfiguration phase, a robust backbone network is formed, encompassing most substations, to facilitate load recovery in the subsequent stage. Finally, in the load restoration phase, large-scale power supply to loads is efficiently carried out [3].

Current schemes of enhancing power system restoration strategies can be divided into two categories: (i) improving entire effective restoration procedure [4]-[5], and (ii) improving efficiency of single stage [6]-[14]. For novel restoration strategies, [4] proposes a restoration strategy based on "generic restoration milestones (GRMs)" concept to improve the system resilience, and a two-stage adaptive restoration decision support system is developed in [5]. For single stage investigation, the traditional three restoration stages are improved in different papers. In black start stage, the generator start-up sequence (GSUS) determines its performance, therefore, [6] provided a new formulation of GSUS as a mixed integer linear programming problem. [7] used the fuzzy analytic hierarchy process for fast restoration of power system after blackouts. [8] formulated the GSUS problem as a preference multi-objective optimization model to realize fast restoration. By using model predictive control, [9] proposed a GSUS strategy for microgrid. In the other two subsequent stages, self-healing algorithm [10], automatic restoration path selection approach [11], and optimal skeleton network reconfiguration [12] are proposed respectively to improve the network configuration, while dynamic frequency response [13] and monitoring system [14] are considered in load restoration stage.

Hydrogen is recognized as a clean energy source and is pivotal to the development of renewable grids with minimal or zero carbon emissions [15]-[16]. The consumption of hydrogen, mainly through fuel cells producing only water and usable heat as byproducts, making it an exceptionally clean energy solution. Therefore, hydrogen-integrated renewable grid with fuel cells is the trend of future power system. However, the black start ability of this system is not fully utilized and investigated for now. Many studies focus on utilizing batteries for black start [17]-[18]. For the hydrogen-integrated grids, an operation strategy in [19] focus on the proactive scheduling of hydrogen facilities, particularly for the distribution network to enhance the grid reliability. Few existing studies utilize fuel cells as black-start resources to facilitate the black start process for hydrogen-integrated system.

To fill this gap, this paper delves into the use of fuel cells and battery energy storage systems as black start resources for power system restoration. First, detailed models that include the output characteristics of fuel cells and battery and the necessary constraints are presented for using them in black start scenarios. Then, the GSUS for systems utilizing fuel cells and batteries are compared, and the efficiency and performance improvements of using fuel cells over traditional methods are highlighted. The in-depth analysis of the optimal location selection for these black start resources and simulations on the IEEE 39-bus system demonstrate that the proposed black start strategy can significantly improve the startup time of generators and overall restoration performance. Sensitivity analyses are conducted to understand the impact of fuel cell capacity and battery storage on restoration efficiency.

## II. BLACK START STRATEGY FOR HYDROGEN-INTEGRATED RENEWABLE GRIDS WITH ENERGY STORAGE

### A. Black start model considering fuel cells as BS resources

During a blackout, generators may shut down and transmission lines may become deenergized, resulting in significant load unserved. A resilient power system should quickly recover from a total blackout and restore service to all loads. The first critical step is to start all shutdown power plants as soon as possible, as restoring generation capability is essential for energizing the transmission infrastructure and facilitating further load pickup.

For the black start stage, the objective is to restore all the generation capability as soon as possible, which is given by:

$$\min \sum_{g \in G} \sum_{t \in} (p_g^{max} - p_g^{start}) \cdot t_g^{start}$$
$$-\beta \cdot \sum_{b \in B} \sum_{t \in T} \frac{(u_{b,t}^B \cdot D_b)}{t} \quad (1)$$

where $p_g^{max}$ is the maximum generation capacity of a power plant, and $p_g^{start}$ is the cranking power required by the plant. $t_g^{start}$ is the time it takes for the power plant to start up after the blackout. The second term prioritizes the restoration of critical loads on specific buses, with $u_{b,t}^B$ as a binary variable indicating whether bus $b$ is restored at time $t$. Besides, $D_b$ is the bus importance degree of bus $b$, and $\beta$ is the weight coefficient for the second term.

At any given time, the restored generation resources, including fuel cells, should be sufficient for the cranking power required to start up generators. It is described as:

$$\sum_{g} p_{g,t} + \sum_{f} p_{f,t} \geq 0, \forall t \in T \quad (2)$$

The minimum and maximum time limitations for starting up a power plant successfully are described by:

$$t_g^{start} \geq t_g^{Min}, \forall g \in G \quad (3)$$
$$t_g^{start} \leq t_g^{Max}, \forall g \in G \quad (4)$$

We use binary variables $u_{f,t}^{start}$, $u_{f,t}^{on}$, $u_{f,t}^{max}$ to indicate different generation status of the fuel cell $f$. When $u_{f,t}^{start}$ is 1, it indicates the fuel cell is starting up. When $u_{f,t}^{on}$ is 1, it indicates the fuel cell is generating power. When $u_{f,t}^{max}$ is 1, it indicates fuel cell is operating at the high end of its range. In order to model the actual generation power from fuel cells injected into the grid, the multiplication of binary variables will be required. In order to linearize it, three ancillary variables $y_{f,t1,t2}^{start}$, $y_{f,t1,t2}^{on}$, $y_{f,t1,t2}^{max}$ are utilized. Referencing our prior work [12], the definition and constraints for these variables are described as follows:

$$y_{f,t1,t2}^{start} = u_{f,t1}^{start} \cdot u_{f,t2}^{start}, \forall f \in F, t1, t2 \in T \quad (5)$$
$$y_{f,t1,t2}^{start} \geq u_{f,t1}^{start} + u_{f,t2}^{start} - 1, \forall f \in F, t1, t2 \in T \quad (6)$$
$$y_{f,t1,t2}^{start} \leq u_{f,t1}^{start}, \forall f \in F, t1, t2 \in T \quad (7)$$
$$y_{f,t1,t2}^{start} \leq u_{f,t2}^{start}, \forall f \in F, t1, t2 \in T \quad (8)$$
$$y_{f,t1,t2}^{on} = u_{f,t1}^{on} \cdot u_{f,t2}^{on}, \forall f \in F, t1, t2 \in T \quad (9)$$
$$y_{f,t1,t2}^{on} \geq u_{f,t1}^{on} + u_{f,t2}^{on} - 1, \forall f \in F, t1, t2 \in T \quad (10)$$
$$y_{f,t1,t2}^{on} \leq u_{f,t1}^{on}, \forall f \in F, t1, t2 \in T \quad (11)$$
$$y_{f,t1,t2}^{on} \leq u_{f,t2}^{on}, \forall f \in F, t1, t2 \in T \quad (12)$$
$$y_{f,t1,t2}^{max} = u_{f,t1}^{max} \cdot u_{f,t2}^{max}, \forall f \in F, t1, t2 \in T \quad (13)$$
$$y_{f,t1,t2}^{max} \geq u_{f,t1}^{max} + u_{f,t2}^{max} - 1, \forall f \in F, t1, t2 \in T \quad (14)$$
$$y_{f,t1,t2}^{max} \leq u_{f,t1}^{max}, \forall f \in F, t1, t2 \in T \quad (15)$$
$$y_{f,t1,t2}^{max} \leq u_{f,t2}^{max}, \forall f \in F, t1, t2 \in T \quad (16)$$

The operation times at different stages of fuel cell generation are constrained as follows:

$$T - \sum_{t} u_{f,t}^{start} \geq T_t^N \cdot (1 - u_{f,t}^{start}), \forall f \in F, t \in T \quad (17)$$

$$T \cdot u_{f,t2}^{start} - T \cdot u_{f,t2}^{on} - \sum_{t1 \in T} y_{f,t1,t2}^{start} + \sum_{t1 \in T} y_{f,t1,t2}^{on} \leq T_{t2}^N$$
$$, \forall f \in F, t2 \in T \quad (18)$$

$$T - \sum_{t} u_{f,t}^{start} \geq (T_t^N - T_f^{Crank}) \cdot (u_{f,t}^{start} - u_{f,t}^{on})$$
$$, \forall f \in F, t \in T \quad (19)$$

$$T \cdot (u_{f,t2}^{on} - u_{f,t1}^{max}) - \sum_{t1 \in T} y_{f,t1,t2}^{on} + \sum_{t1 \in T} y_{f,t1,t2}^{max}$$
$$\leq (T_t^N - T_g^{Crank}) \cdot (u_{f,t2}^{on} - u_{f,t2}^{max}) \quad \forall f \in F, t2 \in T \quad (20)$$

$$T \cdot (u_{f,t}^{on} - u_{f,t}^{max}) \leq T - \sum_{t1 \in T} u_{f,t1}^{start} + T_f^{crank} + T_f^{ramp}$$
$$, \forall f \in F, t \in T \quad (21)$$

$$T \cdot u_{f,t2}^{max} - \sum_{t1 \in T} y_{f,t1,t2}^{max} \leq (T_t^N - T_f^{Crank} - T_f^{Ramp})$$
$$\cdot u_{f,t2}^{max}, \forall f \in F, t2 \in T \quad (22)$$

Using ancillary variables, the actual power injected into the grid from fuel cells is formulated in Equation (23).

$$-p_f^{start} \cdot u_{f,t2}^{start} - u_{f,t2}^{on}$$
$$+ \left( T_f^{ramp} \cdot T_{t2}^N - T_f^{Crank} - p_f^{start} \right)$$
$$* \left( u_{f,t2}^{on} - u_{f,t2}^{max} \right) + T_f^{ramp} \cdot T \cdot \left( u_{f,t2}^{max} - u_{f,t2}^{on} \right)$$
$$+ T_f^{ramp} \cdot \sum_{t1 \in T} y_{f,t1,t2}^{on} - T_f^{ramp} \cdot \sum_{t1 \in T} y_{f,t1,t2}^{max}$$
$$+ \left( p_f^{max} - p_f^{start} \right) \cdot u_{f,t2}^{max} = p_{f,t2}$$
$$, \forall f \in F, t2 \in T \quad (23)$$

The time durations for different operational stages of the fuel cells are described as follows:

$$u_{f,t}^{start} - u_{f,t+1}^{start} \leq 0, \forall t \in T, f \in F \quad (24)$$

$$u_{f,t}^{Max} - u_{f,t+1}^{Max} \leq 0, \forall t \in T, f \in F \quad (25)$$

$$t_f^{Start} = \sum_{t' \in T} \left( 1 - u_{f,t'}^{Start} \right), \forall f \in F \quad (26)$$

$$\sum_{t' \in T} \left( 1 - u_{f,t'}^{On} \right) - t_f^{Start} - t_f^{Crank} \geq 0, \forall f \in F \quad (27)$$

$$\sum_{t' \in T} \left( 1 - u_{f,t'}^{Max} \right) - t_f^{Start} - t_f^{Ramp} - t_f^{Crank} \geq 0, \forall f \in F \quad (28)$$

When the transmission infrastructure is deenergized, it can be restored once its connected transmission lines or substations are energized. The energization of the transmission paths is constrained as follows:

$$u_{g,t}^{start} + u_{f,t}^{start} \leq u_{b,t}^B, \forall t \in T, b \in B, g \in G_b, f \in F_b \quad (29)$$

$$u_{k,t} - u_{b,t}^B \leq 0, \forall t \in T, g \in G, b \in B_k^F \quad (30)$$

$$u_{k,t} - u_{b,t}^B \leq 0, \forall t \in T, g \in G, b \in B_k^T \quad (31)$$

$$u_{k,t} - u_{k,t+1}^B \leq 0, \forall t \in T, k \in K \quad (32)$$

$$u_{k,t+1} - u_{b1,t}^B - u_{b2,t}^B \leq 0$$
$$, \forall t \in T, k \in K, b1 \in B_k^F, b1 \in B_k^T \quad (33)$$

$$u_{b,t}^B - \sum_{k \in K(b)} u_{k,t} - \sum_{g \in G(b)} u_{g,t} - \sum_{f \in F(b)} u_{f,t}^{On} \leq 0$$
$$, \forall b \in B, t \in T \quad (34)$$

In the event of a complete blackout in the area, we assume that all generators, transmission lines, and substations are deenergized, and can be expressed as follows:

$$u_{g,1}^{start} = 0, \forall g \in G \quad (35)$$

$$u_{f,1}^{start} = 0, \forall f \in F \quad (36)$$

$$u_{b,1}^B = 0, \forall b \in B \quad (37)$$

$$u_{k,1} = 0, \forall k \in K \quad (38)$$

The black start generators and fuel cells, which can self-start, are described by:

$$u_{g,2}^{start} = 1, \forall g \in G^S \quad (39)$$

$$u_{f,2}^{start} = 1, \forall f \in F \quad (40)$$

### B. Black start model considering battery as BS resources

For comparison of the performance of fuel cells as black start resources, a black start model considering battery as BS resources is also proposed. The binary variables $w_{b,t}^S$ and $w_{b,t}^E$ are introduced to indicate the start and end of the discharging status of the battery. These variables are constrained as follows:

$$w_{b,t}^S \geq w_{b,t}^E, \forall b \in B^A, t \in T \quad (41)$$

$$w_{b,t}^S \geq w_{b,t-1}^S, \forall b \in B^A, t \in T, t \geq 2 \quad (42)$$

$$w_{b,t}^E \geq w_{b,t-1}^E, \forall b \in B^A, t \in T, t \geq 2 \quad (43)$$

The time $t_b^S$ at which the battery starts discharging after the blackout can be calculated using the $w_{b,t}^S$, which is described as:

$$t_b^S = \sum_t \left( 1 - w_{b,t}^S \right), \forall b \in B^A \quad (44)$$

The battery starting time is constrained by:

$$t_b^S \geq t_b^{Min}, \forall b \in B^A \quad (45)$$

The maximum and minimum discharging power of the battery is described by:

$$\left( w_{b,t}^S - w_{b,t}^E \right) \cdot P_b^{Min} \leq p_{b,t}^{BA} \leq \left( w_{b,t}^S - w_{b,t}^E \right) \cdot P_b^{Max}$$
$$, \forall b \in B^A, t \in T \quad (46)$$

When the battery is discharging, $w_{b,t}^S - w_{b,t}^E$ will be 1, and the discharging power $p_{b,t}^{BA}$ will be constrained by $P_b^{Min}$ and $P_b^{Max}$. When battery is not discharging, $p_{b,t}^{BA}$ will be 0.

The current battery SOC should be greater than the minimum SOC limits, which is described by:

$$SOC_b^S - \sum_t p_{b,t}^{BA} \cdot \frac{T^M}{60} \geq SOC_b^{Min}, \forall b \in B^A \quad (47)$$

In this case, the bus can also be energized when the connected battery is discharging, as described by:

$$u_{b,t}^B - \sum_{k \in K(b)} u_{k,t} - \sum_{g \in G(b)} u_{g,t} - \sum_{b \in BA(b)} w_{b,t}^S \leq 0$$
$$, \forall b \in B, t \in T \quad (48)$$

Additionally, the black start power is provided by the restored generators and batteries, as described by:

$$\sum_{g \in G} p_{g,t} + \sum_{b \in BA} p_{b,t}^{BA} \geq 0, \forall t \in T \quad (49)$$

## III. CASE STUDIES

The proposed black start with fuel cells (BS-FC) and black start with batteries (BS-BT) models are tested, along with the benchmark black start model on the grid without fuel cells and batteries, using the modified IEEE 39-bus system from [12], [20]. This test case contains ten generators and is suitable for the study of black start especially for the evaluation of the generator start up sequence methods. Generator 1 on bus 30 is the only black start generator in the system.

In the simulation, the cranking time of the generators is set to one hour. In the first comparison, two fuel cells or batteries are located on bus 6 and bus 16, which have the highest bus degree importance. Both fuel cells and batteries have 50MW power output capability. The batteries have 50MWh storage with a minimum output of 10% of their maximum output capability. The simulation results are shown in Table I. The systemwide restored power during the black start is illustrated in Fig. 1. It shows that both BS-FC and BS-BT cases, of which curves overlapped with each other in Fig. 1, can significantly improve the startup time of the generators. Compared to the case without fuel cells and batteries, the FC and BT cases provide restoration resources at these two locations, aiding in the restoration of the transmission network and providing initial power for generator startup. The FC case performs better than the BT case due to the limitations of battery storage.

TABLE I. GENERATOR STARTUP TIME FOR IEEE 39-BUS SYSTEM WITH DIFFERENT BLACK START RESOURCES

| Generator number | Without Fuel Cells and Batteries | Fuel Cells (50MW) | Batteries (50MW) |
|---|---|---|---|
| Generator 2 | 140 min | 40 min | 40 min |
| Generator 3 | 220 min | 120 min | 120 min |
| Generator 4 | 160 min | 60 min | 60 min |
| Generator 5 | 240 min | 140 min | 160 min |
| Generator 6 | 180 min | 80 min | 80 min |
| Generator 7 | 200 min | 100 min | 100 min |
| Generator 8 | 80 min | 140 min | 140 min |
| Generator 9 | 220 min | 120 min | 120 min |
| Generator 10 | 200 min | 100 min | 100 min |
| System Average | 182.2 min | 100 min | 102.2 min |

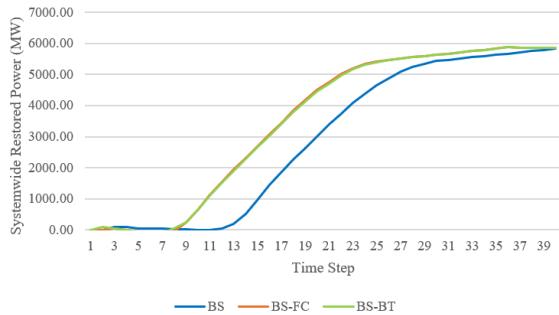

Fig. 1. Systemwide restored power of IEEE 39-bus system during the black start.

TABLE II. GENERATOR START-UP TIME (MINUTES) FOR IEEE 39-BUS SYSTEM WITH DIFFERENT FUEL CELLS GENERATION CAPACITY

| Fuel Cell Capacity (MW) | 100 | 50 | 40 | 30 | 20 | 15 | 10 | 5 |
|---|---|---|---|---|---|---|---|---|
| Gen 3 | 80 | 120 | 80 | 120 | 120 | 140 | 120 | 120 |
| Gen 5 | 140 | 140 | 140 | 140 | 160 | 160 | 160 | 160 |
| Gen 6 | 80 | 80 | 120 | 80 | 100 | 100 | 100 | 140 |
| Gen 7 | 80 | 100 | 140 | 140 | 140 | 100 | 100 | 140 |
| Gen 8 | 120 | 140 | 140 | 140 | 140 | 140 | 160 | 160 |
| Average | 91.1 | 100 | 104.4 | 104.4 | 108.8 | 108.8 | 111.1 | 115.5 |

We also investigate how the capacity of fuel cells influences restoration performance. Since fuel cells can obtain hydrogen from hydrogen pipelines, they do not face storage limits or energy capacity limits like batteries. As shown in Table II, when the output capacity is reduced, some generators experience startup delays. However, not all generators are delayed; some may start earlier. This is because the strategy optimizes the overall system generation. The systemwide restored power for the black start with different fuel cell capacities is plotted in Fig. 2.

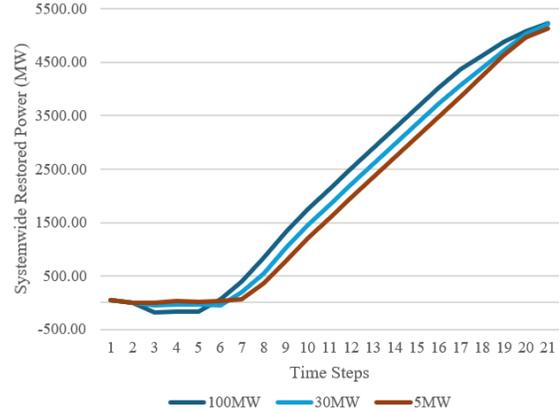

Fig. 2. Systemwide restored power of IEEE 39-bus system with different fuel cell capacities.

We further investigate how the size of batteries influences restoration performance. We assume the minimum output is about 10% of the maximum output capacity for batteries, and the batteries are initially charged to release energy at maximum output power for one hour. The numerical results are shown in Table III.

TABLE III. GENERATOR START-UP TIME (MINUTES) FOR IEEE 39-BUS SYSTEM WITH DIFFERENT BATTERY DISCHARGE CAPACITY

| Discharge Capacity (MW) | 120 | 100 | 80 | 50 | 30 | 20 | 10 | 5 |
|---|---|---|---|---|---|---|---|---|
| Gen3 | 100 | 80 | 120 | 120 | 140 | 120 | 120 | 120 |
| Gen5 | 120 | 140 | 140 | 160 | 140 | 160 | 160 | 160 |
| Gen6 | 80 | 80 | 80 | 80 | 80 | 100 | 100 | 140 |
| Gen7 | 80 | 80 | 80 | 100 | 120 | 140 | 100 | 140 |
| Gen8 | 80 | 120 | 100 | 140 | 120 | 140 | 160 | 160 |
| Gen9 | 120 | 120 | 120 | 120 | 140 | 120 | 120 | 120 |
| Gen10 | 100 | 100 | 100 | 100 | 100 | 100 | 140 | 100 |
| Average | 86.6 | 91.1 | 93.3 | 102.2 | 106.6 | 108.8 | 111.1 | 115.5 |

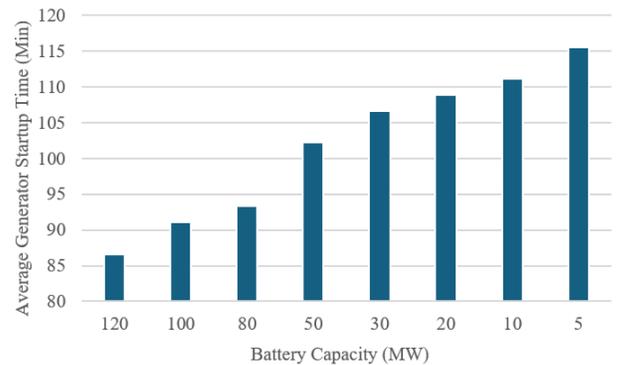

Fig. 3. Average generator startup time for IEEE 39-bus system with different battery capacities.

Comparing fuel cells and battery storage, their output capacity or size similarly influences restoration performance. However, due to storage limitations, a system with batteries may have different GSUS results.

When a blackout occurs, the battery may not be fully charged. We investigated the influence of the initial state of charge (SOC) on restoration performance. Assuming a maximum storage capacity allowing a two-hour release at maximum output power, with a maximum power output of 30 MW, we observed that when SOC is greater than 70%, the black start maintains high performance. However, when SOC is low at the time of the black start, performance is significantly impacted. The numerical results are shown in Table IV.

TABLE IV. GENERATOR START-UP TIME (MIN) FOR IEEE 39-BUS SYSTEM WITH DIFFERENT BATTERY SOC

| Battery SOC | ≥70% | 50% | 30% | 10% |
|---|---|---|---|---|
| Generator 3 | 120 | 140 | 140 | 120 |
| Generator 5 | 140 | 140 | 160 | 160 |
| Generator 6 | 80 | 80 | 100 | 140 |
| Generator 7 | 140 | 120 | 120 | 140 |
| Generator 8 | 140 | 120 | 120 | 160 |
| Generator 9 | 120 | 140 | 160 | 120 |
| System Average | 104.4 | 104.4 | 111.1 | 115.5 |

We also investigated the selection of locations for fuel cells and batteries. In the first selection, the fuel cells and batteries are located on buses with the highest bus importance degree, as used in the simulations above. For the second selection, they are located on buses closer to the generators, within three transmission lines of distance. In this test case, bus 11 and bus 19 are selected. With a fuel cell and battery output capacity of 30 MW, the results are shown in Table V. Based on the results, we can observe that different locations for fuel cells or batteries can significantly impact the generator start-up sequence.

TABLE V. GENERATOR START-UP TIME (MIN) FOR IEEE 39-BUS SYSTEM WITH DIFFERENT BS RESOURCES LOCATIONS

| Configuration | FC | | BT | |
| | Selection 1 | Selection 2 | Selection 1 | Selection 2 |
|---|---|---|---|---|
| Generator 3 | 120 | 140 | 140 | 140 |
| Generator 4 | 60 | 40 | 60 | 40 |
| Generator 5 | 140 | 160 | 140 | 160 |
| Generator 6 | 80 | 140 | 80 | 140 |
| Generator 7 | 140 | 140 | 120 | 140 |
| Generator 8 | 140 | 80 | 120 | 80 |
| Generator 9 | 120 | 120 | 140 | 120 |
| Generator 10 | 100 | 120 | 100 | 120 |
| System Average | 104.4 | 111.1 | 104.4 | 111.1 |

TABLE VI. STATISTICAL DATA FOR BLACK START ON TX-123BT

| Total Number of Large-Scale Thermal Power Plants | 138 |
|---|---|
| Average Time Steps that Generator Start Ramping Up | 5.63 |
| Total Restored Power (MW) | 69,152 |
| Total Number of Branches | 255 |
| Total Number of Buses | 123 |
| Total Number of Critical Branch in BS | 75 |
| Total Number of Critical Bus in BS | 82 |

The black start simulation is also conducted on our created Texas test system TX-123BT [21], to validate the scalability of the proposed black start model. The statistical data for utilizing fuel cells in the black start process for TX-123BT is presented in Table VI. Notably, only 75 out of the 255 total branches are required to restore large-scale thermal power plants.

We also conducted the black start on TX-123BT without the installation of fuel cells. Fig. 4 compares the systemwide restored power during the black start for scenarios with and without fuel cells. We observe that fuel cells can significantly accelerate the black start process. Fig. 5 shows the number of generators ramping up during the black start.

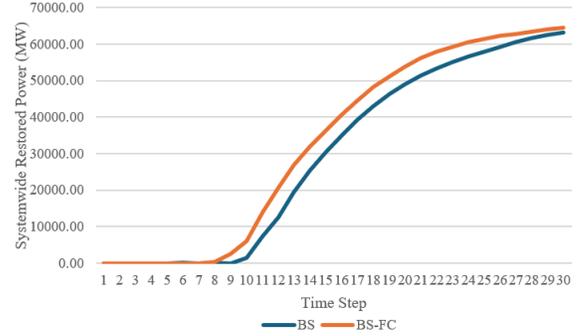

Fig. 4. Systemwide restored power during the black start on TX-123BT.

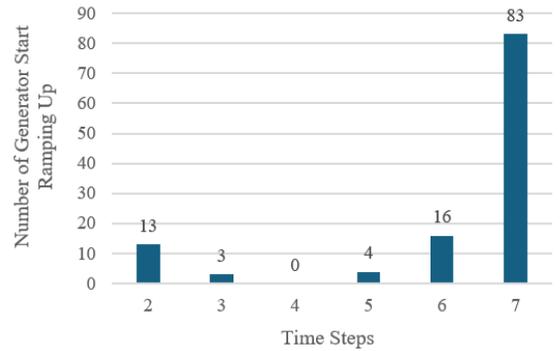

Fig. 5. The number of generators that start ramping up in TX-123BT.

The critical transmission network for black start comprises seven separate areas, with only a portion of the entire network needing restoration. Therefore, network reconfiguration is necessary to identify and reestablish a large and efficient network to support subsequent large-scale load pickup.

## IV. CONCLUSION

This study developed and applied black start models for power systems incorporating fuel cells and battery storage, highlighting their potential as crucial black start resources. The models provide a comprehensive approach to grid restoration, covering energization paths and transmission network energization time, by meticulously integrating the distinct characteristics of fuel cells, such as their power output, and batteries, including their energy storage capacity. Simulations on the IEEE 39-bus system demonstrated distinct generator startup sequences (GSUS) for systems utilizing fuel cells versus batteries. Furthermore, sensitivity analyses on fuel cell capacity, battery storage capacity, initial SOC, and resource locations proved invaluable in identifying the most suitable scenarios for leveraging these clean technologies in black start operations, thereby enhancing grid resilience and mitigating associated challenges.